\documentstyle[aps,prd,twocolumn,epsf]{revtex}
\newcommand \bea{\begin{eqnarray}}
\newcommand \eea{\end{eqnarray}}
\newcommand \ga{\raisebox{-.5ex}{$\stackrel{>}{\sim}$}}
\newcommand \la{\raisebox{-.5ex}{$\stackrel{<}{\sim}$}}

\begin{document}
\title{Fermi systems with long scattering lengths}
\author{Henning Heiselberg}
\address{NORDITA, Blegdamsvej 17, DK-2100 Copenhagen \O, Denmark}

\maketitle

\begin{abstract}
Ground state energies and superfluid gaps are calculated for
degenerate Fermi systems interacting via long attractive scattering
lengths such as cold atomic gases, neutron and nuclear matter.  In the
intermediate region of densities, where the interparticle spacing
$(\sim 1/k_F)$ is longer than the range of the interaction but shorter
than the scattering length, the superfluid gaps and the energy per
particle are found to be proportional to the Fermi energy 
and thus differs from
the dilute and high density limits. The attractive potential increase
linearly with the spin-isospin or hyperspin statistical factor such
that, e.g., symmetric nuclear matter undergoes spinodal decomposition
and collapses whereas neutron matter and Fermionic atomic gases with
two hyperspin states are mechanically {\it stable} in the intermediate
density region. The regions of spinodal instabilities in the 
resulting phase diagram are reduced and do not prevent a superfluid
transition.
\end{abstract}

\vspace{1cm}



\section{Introduction}
Dilute degenerate Fermi systems with long scattering lengths are of
interest for nuclear and neutron star matter (see, e.g.,
\cite{Vijay}). Recently, also dilute systems of cold Fermionic atoms
have been trapped \cite{JILA}.  The number density is sufficient for
degeneracy to be observed and superfluidity is expected at critical
temperatures similar to the onset of Bose-Einstein condensates,
$\sim10-100$nK. S-wave scattering lengths can be very long,
e.g.\footnote{The sign convention of negative scattering length for
attractive potentials is used. Also $\hbar=c=1$.}  $a=-2160$ Bohr
radii for triplet $^{6}Li$ and $a\sim -18.8$~fm for neutron-neutron
interactions. These scattering lengths $|a|$ are much longer than
typical range of the potentials, respectively $R\sim1$~fm for strong
interactions and $R\sim 10-100$~\AA\ for van der Waals forces.
Generally, when $|a|\gg R$ three density regimes naturally emerges:
the low density (or dilute, $k_F^{-1}\ga |a|$), the high density
($k_F^{-1}\la R$), and the {\it intermediate density region} ($R\la
k_F^{-1}\la |a|$).

The latter ``novel'' region of densities is the object of study here.
It will be shown that previous conjectures based on extrapolations
from the dilute limit fail. Instead it is found that both the energy
per particle and superfluid gaps scale with the Fermi energy. They
depend only on statistical factors but not on the scattering length,
range or other details of the interaction. Consequently, phase
diagrams are dramatically altered and the stability criteria differ so
that two spin systems are actually {\it stable}.

The manuscript is organized as follows.  In Sec. II the scaling and
poles of the effective scattering amplitude are studied in a
homogeneous many-body system going from dilute to intermediate
densities. In Sec. III the ground state energies are calculated at
intermediate densities and compared to well known results from the
dilute and high density limits.  In Sec. IV extensions to finite
temperatures are discussed and a phase diagram is constructed
displaying regions of superfluidity and spinodal instabilities.  In
Sec. V the properties of finite systems of Fermions are investigated
as they are relevant for current experiments with magnetically trapped
cold atoms.  Finally, a summary and conclusion is given.

\section{The effective scattering amplitude and superfluidity}
Consider a homogeneous many-body system of fermions of mass $m$ at
density: $\rho=\nu k_F^3/6\pi^2$, where $\nu$ is the statistical factor,
e.g. $\nu=4$ for symmetric nuclear matter and $\nu=2$ for neutron matter
as well as for the $^{40}K$ atomic gas with two hyperspin states
currently studied at JILA \cite{JILA}. The scattering lengths and
Fermi momentum $k_F$ are assumed the same for all
spin-isospin/hyperspin components in the system but interesting
effects of varying the relative densities of the various components
will be discussed at the end. Particles are assumed non-relativistic
and to interact through attractive two-body contact interactions. The
details of the potential is not important, only its range $\sim R$ and
scattering length $a$. We shall be particular interested in cases
where $R\ll|a|$, which occur when, for example, the two-body
potential can almost support a bound state or resonance.

At dilute or intermediate densities the particles interact
via short range interactions that appear singular on length scales
of order the interparticle distance $\sim k_F^{-1}$.
Such systems are best described by resumming the multiple interactions
in terms of the scattering amplitude. The
Galitskii's integral equations \cite{Galitskii} for the effective
two-particle interaction or
scattering amplitude in the medium is given by the ladder resummation
\bea
  \Gamma({\bf p,p',P}) &=& \Gamma_0({\bf p,p',P})
               + m\sum_{\bf k} \Gamma_0({\bf p,k,P})  \nonumber\\ 
  &\times&\left[\frac{N({\bf P,k})}{\kappa^2-k^2} - \frac{1}{\kappa^2-k^2}
  \right]  \Gamma({\bf k,p',P})   \, . \label{Galitskii}
\eea
Here, $\Gamma_0=4\pi a/m$ is the s-wave scattering amplitude in
vacuum; the total energy of the pair in the center of mass is $\kappa^2/m$;
${\bf p,k,p'}$ are the relative momentum of the
pair of interacting particles in the initial, intermediate and final states
respectively, and ${\bf P}$ the total momentum; 
$N({\bf P,k})=+1$ for particle-particle propagation ($|{\bf
P}\pm{\bf k}|\ge k_F$), $N({\bf P,k})=-1$ for hole-hole propagation ($|{\bf
P}\pm{\bf k}|\le k_F$), and zero otherwise. For spin independent
interactions the amplitudes contain a factor $(1-\delta_{\nu_1\nu_2})$
that takes exchange into account between identical spins $\nu_1=\nu_2$.

For obtaining BCS gaps it is sufficient to study pairs with ${\bf P}=0$ where
Eq. (\ref{Galitskii}) is simply
\bea
  \Gamma=\Gamma_0 + 2m\sum_{k\le k_F} \Gamma_0 \frac{1}{\kappa^2-k^2} \Gamma
  \,. \label{Cooper}
\eea
Note the factor of 2 due to particle-particle and hole-hole 
propagation each contributing by the same amount in non-dense systems.
The ladder resummation implicit in Eq. (\ref{Galitskii}) insures that
only  momenta smaller than Fermi momenta enter.
$\Gamma_0$ vary on momentum scales $\sim 1/R\gg k_F$ only and
can therefore be considered constant at low and intermediate densities.
Eq. (\ref{Cooper}) is then easily solved for momenta near
the Fermi surface
\bea
  \Gamma &=& \Gamma_0\left[1-\frac{2}{\pi}k_Fa
  (2+\frac{\kappa}{k_F}\ln\frac{k_F-\kappa}{k_F+\kappa})\right]^{-1}
 \,. \label{G3}
\eea
The in-medium scattering amplitude has a pole due to Cooper pairing when
\bea
   \Delta &\equiv& \frac{k_F^2-\kappa^2}{m} 
   = \frac{(k_F+\kappa)^2}{m}\exp(\frac{\pi}{2\kappa a}-2\frac{k_F}{\kappa})
   \nonumber\\
   &\simeq& E_F \frac{8}{e^2} \exp(\frac{\pi}{2k_Fa}) ,\quad k_F|a|\ll 1\,,
     \label{gap}
\eea
where $E_F=k_F^2/2m$ is the Fermi energy. The critical temperature is
$T_c=(\gamma/\pi)\Delta$, where $\gamma=e^C$ and $C=0.577$ is 
Euler's constant.
Eq. (\ref{gap}) is the BCS gap in the dilute limit which agrees with
gaps calculated in \cite{Melo}.

However, Gorkov et al. \cite{Gorkov} showed that spin fluctuations lead to
a higher order correction $\sim (k_Fa)^2$ in the denominator of 
Eq. (\ref{G3}) that
is amplified by logarithmic terms $\sim \ln(\Delta)\sim 1/k_Fa$.
It contributes by a (negative) constant in the exponent and leads to a 
reduction of the gap in the dilute limit
by a factor $(4e)^{1/3}=2.215...$ as compared to Eq. (\ref{gap})
for two spins. Generally for $\nu$ spins, isospins or hyperspins
the gap is \cite{gap}
\bea
   \Delta = E_F \frac{8}{e^2} (4e)^{\nu/3-1} 
   \exp\left[\frac{\pi}{2ak_F}\right] \, .  \label{Gorkov}
\eea

 In the intermediate density region pairing must still occur since
interaction are attractive. The validity of the expressions of
Eqs. (\ref{gap},\ref{Gorkov}) in this density regime will be discussed
further below. They predict that in the limits $a\to-\infty$ and $R\to
0$ the gap cannot depend on either $a$, $R$ or other details of the
potential.  For dimensional reasons the gap can therefore only be
proportional to the Fermi energy.

\section{Ground state energies}
The ground state energy is another crucial property of 
the system. In terms of the on-shell
effective scattering amplitude it is \cite{Galitskii,FW}
\begin{eqnarray}
  E &=& \sum_{k_1\nu_1} \frac{k_1^2}{2m}  + \frac{1}{2}
  \sum_{k_1k_2\nu_1\nu_2} \Gamma({\bf p,p,P})\, (1-\delta_{\nu_1\nu_2})
     \nonumber \\
              &=& \frac{3}{5} \frac{k_F^2}{2m} N
     +\frac{\nu(\nu-1)}{2}\sum_{k_1k_2} \Gamma({\bf p,p,P}) 
        \,. \label{EV}
\end{eqnarray}
Here,
$N=V\rho$ is the number of particles and
the summations $\nu_1,\nu_2$ include spin and isospin or hyperspin states.
The factor $(1-\delta_{\nu_1\nu_2})$ in the amplitude
due to exchange has now been written explicitly.
Antisymmetrization of the wave-function prevents identical particles to
be in relative s-states.
At low and intermediate densities, $k_FR\ll 1$, 
the exchange term is $1/\nu$ of the direct one for spin independent
interactions.

Before investigating the novel intermediate density region,
a brief review of results at low and high densities is given.

\subsection{Low densities (dilute): $k_F|a|\ll 1$}
At low densities, $k_F|a|\ll 1$, gaps are small and have little effect
on the total energy of the system. Expanding the effective scattering
amplitude of Eq. (\ref{Galitskii})
in the small quantity $k_Fa$, the energy per particle
is obtained from Eq. (\ref{EV}) by summing over momenta of the two
interacting particles
\bea
   \frac{E}{N} &=& E_F [\frac{3}{5}+(\nu-1)\frac{2}{3\pi} k_Fa
     \nonumber\\
  &+&(\nu-1)\frac{4(11-2\ln2)}{35\pi^2}(k_Fa)^2
     + {\cal O}((k_Fa)^3) ]  
  \,.\label{Lenz}
\eea
It consists of respectively the Fermi kinetic energy, the standard
dilute pseudo-potential \cite{Lenz} proportional to the scattering length
and density, and orders $(k_Fa)^2$ \cite{HY} and higher \cite{Efimov}. 

The zero temperature hydrodynamic sound speed squared can
at low temperatures be expressed as
\bea
  s^2 =\frac{1}{m} \left(\frac{\partial P}{\partial \rho}\right)
      =\frac{1}{m} \frac{\partial}{\partial\rho}  \left(\rho^2
      \frac{\partial E/N}{\partial\rho}\right) \,,\label{s}
\eea
With the energy per particle of Eq. (\ref{Lenz}) 
at low densities, the sound speed can be expanded as
\bea
  s^2=\frac{1}{3} v_F^2 \left(1+\frac{2}{\pi}(\nu-1)k_F a +...\right) 
      \,.\label{sl}
\eea
where $v_F=k_F/m$ is the Fermi velocity.
It is commonly conjectured from the first two orders that the
Fermion (and Bose) gases undergo spinodal
instability when the sound speed squared becomes negative,
which occurs when $k_Fa\la -\pi/2(\nu-1)$.
However, at the same densities the dilute approximation leading
to Eq. (\ref{Lenz}) fails and so does the conjecture 
as will be shown below.

\subsection{High densities, $k_F R\gg 1$}
At high densities, $k_FR\ga 1$, the particle potentials overlap
and each particle experience on average the volume integral of the
potentials.
The energy per nucleon consists of the Fermi kinetic energy and the 
Hartree-Fock potential, (see, e.g. \cite{FW} Eq. 40.14):
\begin{eqnarray}
   \frac{E_{HF}}{N} &=& \frac{3}{5} E_F
     +\frac{\rho}{2} \int d^3r\,V(r)\left[ 1
 -\frac{1}{\nu}\left(\frac{3j_1(rk_F)}{rk_F}\right)^2 \right]
   .  \label{Hartree}
\end{eqnarray}
The latter term is the exchange energy which vanishes at very high
densities,
$k_FR\sim R/r_0\gg 1$, leaving the Hartree potential term only.
At lower densities, $k_FR\sim R/r_0\ll 1$,
it is identical to the first integral, i.e., the Hartree direct term is
$\nu$ times the Fock exchange term as also found in the dilute limit,
Eqs. (\ref{Ei}-\ref{Lenz}).

As shown in \cite{PPT}, the Hartree potential is considerably less
attractive than the dilute potential. In fact it vanishes when the range
of the interaction goes to zero and the scattering length to infinity.
Take for example a square well potential of range $R$ and depth
$-V_0$. Long scattering lengths requires $V_0R^2\to\pi^2/4m$, and
therefore the Hartree potential is $\propto V_0R^3\sim R\to 0$.
Only in the Born approximation do the Hartree (\ref{Hartree})
and dilute potentials (\ref{Lenz}) coincide since the
Born scattering length is $a_{Born}=(m/4\pi)\int d^3r V(r)$.

Short range repulsion complicate the high density limit. In nuclear
and atomic system the repulsive core is only of slightly shorter range
than the attractive force. It makes the liquid strongly correlated
and the Hartree-Fock approximation fails \cite{BBG,FW}.
How the short range repulsion turn the attraction to repulsion at
these even higher densities will, however, not affect the
intermediate density region.

\subsection{Intermediate densities, $|a|\gg k_F^{-1}\gg R$}
At intermediate densities, $|a|\gg k_F^{-1}\gg R$, the scattering
length expansion in Eq. (\ref{Lenz}) breaks down.  Brueckner, Bethe
and Goldstone \cite{BBG} pioneered such studies for nuclear matter and
$^3He$ where the range of interactions, scattering lengths and
repulsive cores all are comparable in magnitude.  In our case the
range of interaction is small, $k_FR\ll1$, and therefore all
particle-hole diagrams are negligible. Higher order particle-particle
and hole-hole diagrams do contribute by orders of $\sim
\Gamma_0(mk_F\Gamma)^n$, It is evident from Eq. (\ref{gap}) that at
intermediate densities $\Gamma$ no longer is proportional to
$\Gamma_0$ or the scattering length but instead $\Gamma\propto
(mk_F)^{-1}$. Due to the very restricted phase space such higher
order terms are usually neglected as in standard Brueckner theory.
$\Gamma$ of Eq. (\ref{Galitskii}) can therefore be considered as a
resummation of an important class of diagrams.
The Cooper instability complicates the calculation of $\Gamma$. If
the gap is small the instability occurs only for pairs near the Fermi
system with opposite momenta and spin and the effect on the total energy
is small. The momentum dependence of the effective scattering
amplitude also complicates a self-consistent calculation. These
complications can be dealt with by approximating $\Gamma$ by its momentum
average value in Eq. (\ref{Galitskii}). The momentum integrals
are then analogous to those in the dilute limit (\ref{Lenz}), 
and one obtains from Eq. (\ref{EV}) 
\bea
 \frac{E}{N} \simeq E_F \left[ \frac{3}{5} + \frac{(\nu-1)\frac{2}{3\pi} k_Fa}
 {1-\frac{6}{35\pi}(11-2\ln2)k_Fa} \right] \,. \label{Ei}
\eea
This expression is valid 
for dilute systems, where it reproduces Eq. (\ref{Lenz}),
and approximately valid within the Galitskii ladder resummation
at intermediate densities, $R\ll k_F^{-1}\ll |a|$, where 
it reduces to
\bea
 \frac{E}{N} = E_F \left[ \frac{3}{5} - (\nu-1)c_1\right]
             = E_F c_1(\nu_c-\nu) \,, \label{Ei2}
\eea
with $c_1=35/9(11-2\ln2)\simeq 0.40$ and $\nu_c=1+3/5c_1\simeq
2.5$.  Both the attractive and the kinetic part of the energy per
particle are proportional to the Fermi energy at these intermediate
energies as found for the gaps above.

The other remarkable feature of Eq. (\ref{Ei2}) is that the energy per
particle changes sign for a critical number of degrees of freedom,
$\nu_c\simeq 2.5$.  Fermi systems with more degrees of freedom such as
symmetric nuclear matter have negative energy per particle and will
therefore implode, undergo spinodal decomposition and fragment
\cite{HPR}. Contrarily, systems with $\nu\la\nu_c$ such as neutron matter
and atomic systems with only two hyperspins have positive energy per
particle and will therefore explode, if not contained.  This is also
evident from the sound speed squared which from Eqs. (\ref{s}) and (\ref{Ei2})
becomes 
\bea
   s^2=(5/9)c_1(\nu_c-\nu)v_F^2 \,.
\eea

Calculations for pure neutron matter and symmetric nuclear matter at
low densities by variational Monte Carlo \cite{Vijay} 
and in neutron matter by Pade' approximants to R-matrix calculations
\cite{Baker} independently confirm the above results approximately
in a limited range of intermediate densities.  In the density
range $\rho_0\ga\rho\ga |a|^{-3}\simeq 10^{-4}\rho_0$ the energy per
particle is positive for neutron matter and negative for symmetric
nuclear matter \cite{Vijay}. 
They scale approximately with $\rho^{2/3}$ with
coefficients compatible with Eq. (\ref{Ei2}). In symmetric nuclear
matter the intermediate density regime is, however, limited since
protons also interact through the triplet channel, which has a shorter
repulsive scattering length $a_t\simeq 5.4$~fm, besides the singlet
one, $a_s=-18.8$~fm, relevant for neutron matter.  Never-the-less, the
ladder resummation in the Galitskii integral equation
Eqs. (\ref{Ei},\ref{Ei2}) are supported by dimensional arguments and
quantitatively it successfully predicts $\nu_c$ between that of
neutron and symmetric nuclear matter. The ladder resummation therefore
seems to include the most important class of diagrams.  However, even
small corrections can be important for the magnitude of the
gap because they
appear in the exponent as, e.g., found for induced interaction
(compare Eq. (\ref{Gorkov}) with
(\ref{gap})). In addition, the superfluidity decrease the energy of the
system by $\sim\Delta^2/2E_F$, which can be significant at
intermediate densities if gaps really are as large as the Fermi
energy.

The interesting feature of the intermediate density region, that
energies and gaps are independent of the scattering length, leads
to the remarkable fact that the system is also insensitive to whether
the scattering length goes to plus or minus infinity. In other words,
a many particle system is insensitive to whether the two-body system
has a marginally bound state just above or below threshold;
the two-body bound state or resonance is dissolved in matter at 
sufficiently high density, $k_F|a|\ga 1$. 
For positive scattering lengths a pair condensate of molecules
may form at low densities but they dissolve at intermediate densities
when the range of the two-body wavefunction exceeds the
interparticle distance.

\begin{figure}
\epsfxsize=8.6truecm 
\epsfbox{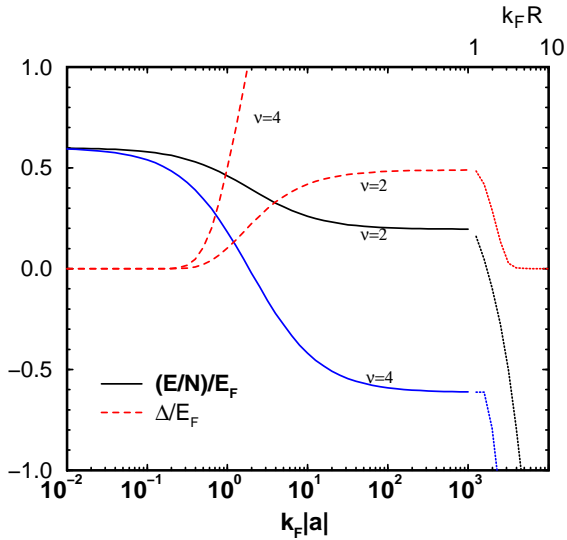}
\caption[]{Ground state energy and superfluid gaps for a degenerate
system of fermions interacting via attractive forces with $R\ll|a|$.
The energy per particle $E/N$ (Eq. (\ref{Ei}), full curves) and the
BCS gap $\Delta$ (Eq. (\ref{Gorkov}), dashed curves) are plotted in
units of the Fermi energy as function of density for $\nu=2$ and
$\nu=4$. Dotted curves to the right show qualitatively the gap and
energy per particle at high density (see text) as a function of $k_FR$
(upper axis). }
\label{fig1}
\end{figure}

In Fig. 1 the energy per particle is shown as function of density
extending from the dilute and intermediate densities,
Eqs. (\ref{Lenz},\ref{Ei}), to high densities, Eq. (\ref{Hartree}).
At low densities the Fermi kinetic energy dominates but at
intermediate densities, $R\la k_F^{-1}\la |a|$, the attractive
potential lower the energy by an amount proportional to the
statistical factor. The two cases $\nu=2$ and $\nu=4$ are seen to
saturate at positive and negative energies respectively.  In the high
density limit the attractive (Hartree) potential of
Eq. (\ref{Hartree}) dominates and will lead to collapse of all Fermi
systems in the absence of repulsive cores. 

 Fig. 1 also shows the superfluid gaps of
Eqs. (\ref{gap},\ref{Gorkov}) for dilute and intermediate density
Fermi systems. At low densities they decrease exponentially as
$\Delta\sim E_F\exp(-2/\pi k_F|a|)$ whereas at intermediate densities
the gaps are a finite fraction of the Fermi energy. At high densities
the gap generally decreases rapidly with density \cite{Emery,FW}.  For
example, for an attractive square well potential of range $R$ and
depth $V_0$ with long scattering length (or marginally bound state),
i.e.  $V_0R^2\simeq \pi^2/4m$, the gap decrease exponentially as
$\Delta\sim \exp(-4k_FR/\pi)$.  When $|a|\gg R$ plateaus appear for
$(E/N)/E_F$ and $\Delta/E_F$. Since $E_F$ also decrease with
decreasing density, the gap itself is narrowly peaked near $k_F\simeq
1/R$ as found in nuclear and neutron matter \cite{Emery}.

Information on the density dependence can be obtained independently
from calculations within the Wigner-Seitz cell approximation that has
recently been employed for the strongly correlated nuclear liquid
\cite{HH}.  The periodic boundary condition is a computational
convenience which contains the important scale for nucleon-nucleon
correlations given by the interparticle spacing.  It naturally gives
the correct low density Eq. (\ref{Lenz}) and high density
Eq. (\ref{Hartree}) limits. At intermediate densities one obtains
\bea
   \frac{E}{N} &=& E_F \left[ \frac{3}{5} 
   -c_2 \frac{\nu-1}{\nu^{1/3}} \right]  \,.
   \label{eWS}
\eea
Finite crystal momenta complicates the calculation of $c_2$. A lower
(but reasonable) estimate $c_2\simeq0.25$ can be calculated.  The
potential energy in Eq. (\ref{eWS}) is also proportional to the
kinetic one as found in Eq. (\ref{Ei2}) and of similar magnitude.  The
scaling with $\nu^{-1/3}$ arise because energies scale with the square
of the inverse particle spacing, $r_0^{-2}$, in the Wigner-Seitz cell
approximation, and $\rho=\nu k_F^3/6\pi^2=(4\pi r_0^3/3)^{-1}$.  The
energy per particle can be calculated at all densities and finite
values of $a$ and $R$ and the cross over between the three density
regimes generally confirm the energy per particle shown in Fig. 1.

\section{Phase diagram}

Constructing a phase diagram from the low temperature degenerate
regime to the high temperature classical regime requires a finite
temperature generalization. For illustration we shall follow the
procedure as in Ref. \cite{Baym} and employ the high temperature
approximation for the additional thermal pressure. At high
temperatures quantal effects are negligible and the energy per
particle is simply given by the classical value $E/N\simeq
3T/2$. 
The isothermal sound speed is within this approximation
\bea
   s^2_T = \frac{T}{m} + s^2_{T=0} \,,\label{sT}
\eea 
where the zero temperature sound speed is given by Eq. (\ref{s}) with
energy per particle from Eqs. (\ref{Ei},\ref{Lenz}).  The spinodal
instability condition, $s_T=0$, determines the line of collapse
$T(\rho)$ for long wavelength density fluctuations.

\begin{figure}
\epsfxsize=8.6truecm 
\epsfbox{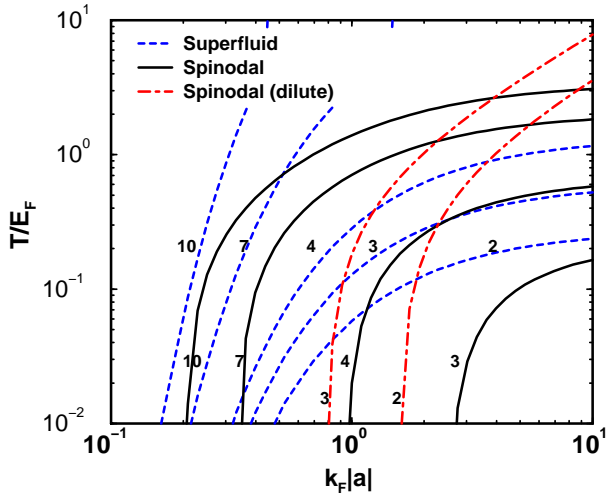}
\caption[]{Phase diagram at low and intermediate densities $(\rho=\nu
k_F^3/6\pi$) for a gas of fermions interacting via a long (attractive)
scattering length $a$. Spinodal lines are shown with full curves and
the superfluid transition by dashed curves for various number of spin
states $\nu$ as labeled.  The area constrained to the lower right
corner are the spinodally unstable and superfluid regions.
As systems with two spin states only are stable at intermediate
densities the $\nu=2$ spinodal line is absent.  Contrarily, the
$\nu=2$ spinodal line based on extrapolating the dilute approximation
to higher densities (see text) is shown by dash-dotted curves.  }
\label{figphase}
\end{figure}

The resulting phase diagram is shown in Fig. 2 for $\nu=2,3,4,7,10$
spin states.  The lower ($T\la T_F$) and right ($k_F|a|\ga1$) corner
of the phase diagram is the spinodally unstable region where the
system collapses. The region decreases for fewer spin states and is
absent for $\nu=2$. For comparison the spinodal lines for $\nu=2$ and $\nu=3$
are shown when the dilute approximation of Eq. (\ref{Lenz}) is
extrapolated into intermediate densities. Generally, the spinodally
unstable regions based on the dilute approximation are substantially
overestimated.

The regions of superfluidity given by $T_c=(\gamma/\pi)\Delta$ and
Eq. (\ref{Gorkov}) are also shown in Fig. 2. As for the spinodally
unstable region it is the lower right corner that is
superfluid. However, superfluidity extends to lower densities and
therefore mechanical instability does not prevent the BCS-type pairing
in the case of fermions. The opposite conclusion was reached for the
pairing transition in Bose-Einstein condensates \cite{Baym}.  As
cooling becomes increasingly difficult at temperatures below the Fermi
temperature we observe that the superfluid transition is readily
obtained by increasing the density above $k_F|a|\ga 1$.

The phase diagram is quantitatively correct at low as well as high
temperatures. Around the Fermi energy it gives a qualitative
description only due to the approximate thermal pressure employed.
Furthermore, at intermediate densities the superfluid gaps become
large exceeding the Fermi energy for large spins, and the
corrections to the ground state energies can therefore not be ignored.

\section{Finite systems}
The degenerate Fermi gases and Bose-Einstein condensates (BEC) 
produced so far contain $n\sim10^3-10^6$ magnetically trapped
alkali atoms. 
Some of them interact via long scattering lengths such as the 
triplet $^6Li$ fermions with $a=-2160$ Bohr radii and singlet $^{85}Rb_2$ 
bosons with $|a|\ga 10^3$ Bohr radii.  
Large scattering lengths $a\to\pm\infty$ can be taylored by using 
Feshback resonances, i.e. hyperfine
states close to threshold further tuned by magnetic fields.
Fermi gases differ from BEC's in several
respects. Most importantly, whereas bosons sit at zero momentum states,
fermions have considerable kinetic energy. Therefore, when
interactions are small, a BEC has energy per particle $\hbar\omega$ and
size $a_{osc}$, where $\omega=(\omega_\perp\omega_z)^{1/3}$ is the
geometric average of the magnetic trap frequencies and
$a_{osc}=\hbar/\sqrt{m\omega}$ is the oscillator length. A degenerate
gas of $N$ Fermionic atoms has a larger energy per particle
$\sim N^{1/3}\hbar\omega$ and
size $L\sim N^{1/6}a_{osc}$.  In current experiments with degenerate
Fermi gases and BEC's the densities are low so that the dilute
potential applies and the energy per particle is approximately
\bea
  \frac{E}{N} \simeq \frac{3}{4}\left(\frac{6}{\nu}\right)^{1/3}
 N^{1/3} \hbar\omega 
 + \frac{\nu-1}{\nu}\frac{2\pi a}{m} \frac{N}{L^3} \,,\label{BEC}
\eea
where the average density in the trap has been approximated by
$\langle\rho\rangle\simeq N/L^3$ \cite{Stringari}.  For a small number
of trapped atoms with attractive scattering lengths the system is
meta-stable but for a large number of trapped atoms, $N\ga
\nu(a_{osc}/(\nu-1)a)^6$, the attractive potential overcomes the 
Fermi kinetic energy and the degenerate Fermi gas
 becomes unstable and implodes.  However, around the same
density $k_F|a|\ga 1$ and we enter the intermediate density region,
where the potential of (\ref{Ei}) should be applied instead of the
dilute potential. The gas is therefore mechanically stable for two
hyperspins only contrary to conclusions based on the dilute potential
\cite{JILA,Houbiers}.

Recent experiment on cold magnetically trapped Fermionic atoms
\cite{JILA} observed egeneracy for $^{40}K$ atoms in the two hyperfine
states $m_F=9/2,7/2$. Current experimental oscillator lengths
$a_{osc}\simeq\mu m$ are less than one order of magnitude longer than the
atomic scattering length $|a|$ of $^6Li$.  It should be possible to
reach intermediate densities, $k_F|a|\ga1$, by trapping $N\ga 10^6$
$^6Li$ atoms \cite{Houbiers}.  The atomic gases offer the unique
opportunity to vary the densities as well as the relative amount of
the hyperfine states.  Varying the composition is a convenient way to
vary the gaps and attractive potential of Eqs.
(\ref{Lenz},\ref{Ei},\ref{Hartree}) through $\nu$ for given density
and scattering length. In the limit where most atoms are in one of the
states, the Fock and Hartree terms almost cancels and effectively
$\nu\to 1_+$.

More intricate systems of mixtures of fermions and bosons, e.g.
$^{39,40,41}K$ isotopes can also be studied.  If the interaction is
attractive it will contract the atomic cloud towards higher densities.
Irrespective of whether the bosons or fermions attract or repel the
induced interactions, which are of second order in the fermion-boson
coupling, enhance the gap \cite{gap}.

An artificial ``gravitational'' or ``Coulomb'' force can be exerted on
the atoms by shining laser light on the trapped cloud from many
directions \cite{Odell}. It would add an energy per particle of order
$\sim GNm^2/L$ to Eq. (\ref{BEC}) where $G$ is proportional to the laser
field intensity. Such an interaction has several interesting
consequences.  If $G$ is attractive, it would contract the cloud
towards higher densities which would increase gaps (see also
\cite{Luciano}) and for sufficiently large $G$ the
intermediate density region is entered. Depending on
the strengths and sign of the scattering amplitude and gravitational
interactions the kinetic energy of the atoms will be balanced by the
magnetic trap and/or the scattering or gravitational interactions.
The resulting phase diagram is much more complex.

If such a strong attractive laser field is suddenly applied to
the gas, the Jeans instability sets in and the gas 
collapses until balanced again by the kinetic energy.
Subsequently, the system will ``bounce'' analogous to the initial stages of
a supernova explosion. If, however, intermediate energies
are reached and the number of spins exceed $\nu\ga2.5$, then
the collapse will be further accelerated by the attraction between
atoms. The corresponding critical particle number is
\bea
   N_c \simeq (Gm^3a)^{-3/2} \,,
\eea
at zero temperature. It differs from the standard Chandrasekhar
mass by a factor $(m|a|)^{3/2}$
because the instability condition is $k_F|a|\simeq 1$ whereas
stars go unstable when the particles become relativistic $k_F\simeq m$.

\section{Summary}
The energy per particle and superfluid gaps have been calculated for
an homogeneous system of fermions interacting via a long attractive
s-wave scattering length.  In the intermediate region of densities, where the
interparticle spacing $(\sim 1/k_F)$ is much longer than the range of
the interaction but much shorter than the scattering length or $|a|$,
the energy per particle and superfluid gaps are proportional to the
Fermi energy.  The energy per particle increases linearly with the
spin-isospin or hyperspin statistical factor such that, e.g.,
symmetric nuclear matter is unstable in the intermediate density
regions and undergoes spinodal decomposition whereas neutron matter
and Fermionic atomic gases with few hyperspin states are mechanically
stable.

 A phase diagram of Fermi gases at low and intermediate densities was
constructed by including thermal pressures in the high temperature
classical approximation.  With the proper energy per particle at
intermediate densities the spinodal region in the phase diagram was
reduced substantially as compared to conjectures based on
extrapolations from the dilute limit.  Generally, mechanical
instability does not prevent a superfluid transition for a wide range
of densities.  This is contrary to Bose gases, where spinodal
instabilities exclude pairing transitions \cite{Baym}.

The interaction energies of the many-body system were discussed for
magnetically trapped cold degenerate gases of Fermi atoms.  In such
systems both superfluidity and the intermediate density region should
be attainable. In these ``novel'' density regions the superfluid gaps
can be large and the stability and sensitivity to the statistical
factor $\nu$ can be studied.  Adding a gravitationally like force by
shining laser light on the atomic cloud further increase densities
whereby collapse and bounce analogous to the early stages of supernova
explosions may be studied.


\end{document}